\documentclass[preprint, showpacs,preprintnumbers,amsmath,amssymb]{revtex4}
\usepackage[dvips]{graphicx}
\usepackage{graphicx}
\usepackage{amsfonts}
\usepackage{bm}
\usepackage{amsmath}
\usepackage{amssymb}
\usepackage{color}
\usepackage[all]{xy}

\def\be{\begin{equation}}
\def\ee{\end{equation}}
\def\bea{\begin{eqnarray}}
\def\eea{\end{eqnarray}}

\begin{document}

\title{GEOMETROTHERMODYNAMICS OF BLACK HOLES IN LORENTZ NON--INVARIANT MASSIVE GRAVITY}

\author{Alberto S\'anchez }
\email{asanchez@ciidet.edu.mx} \affiliation{Departamento de
Posgrado, CIIDET,\\{\it AP752}, Quer\'etaro, QRO 76000, MEXICO}

\date{\today}

\begin{abstract}

We analyze a static and spherically symmetric hairy black hole
solution in non--invariant massive gravity. The formalism of
geometrothermodynamics is used to describe the thermodynamic
characteristics of this black hole in a Legendre invariant way.
For a black hole in massive gravity,  the geometry of the space of
equilibrium states is computed showing that it contains
information about the thermodynamic interaction, critical points
and phase transitions structure.

{\bf Keywords:} Thermodynamics, Phase transitions, Black Holes

\end{abstract}

\pacs{05.70.Ce; 05.70.Fh; 04.70.-s; 04.20.-q}

\maketitle

\section{Introduction}
\label{intro}

The search for a quantum theory of gravity has received attention
of many researchers. These investigations have suggested that
general relativity will be superseded by a quantum theory of
gravity at high enough energies with respect to the Planck mass or
the corresponding Planck length \cite{Donoghue}. The main idea is
that general relativity is valid as an effective field theory for
length scales much larger that Planck length. If this theory is
true, the possibility to proof it by experiments is almost
hopeless. On the other hand, the most successful cosmological
model that is in agreement with the observational data implies the
existence of a vacuum energy related with the cosmological
constant $\Lambda$, whose magnitude is unnatural from the
effective field theory point of view\cite{Copeland}. Hence, a dark
energy is needed to reconcile general relativity with the
observations. The dark energy solves the problem of the
accelerated expansion of the Universe. Nevertheless, the interest
to explain the acceleration of the Universe without resorting to
the dark energy has motivated the research for large--distances
modified theories of the gravity; the Lorentz--breaking massive
gravity is one of these models which is free from pathologies such
as ghosts, low strong coupling scales or instabilities at full
non-perturbative level \cite{Arkani,Arkani2,dubovsky}. Although
these models do not require the existence of $\Lambda$, the
cosmological constant problem remains open. A  generalized
Schwarzschild solution for this model has been obtained by D.
Comelli et al. which is an exact black hole solution  showing a
nonanalytic hair \cite{comelli}.

The study of the thermodynamics of black holes and its
relationship with geometry has been a subject of intensive
research \cite{Bravetti,Aman,Aman2,Aman3,shen,Cai}. This geometric
study has been considered in several papers by means of different
approaches like Weinhold's theory \cite{Weinhold}, Ruppeiner's
theory \cite{Ruppeiner} or the most recent theory called
geometrothermodynamics\cite{quevedo2}. Geometrothermodynamics
(GTD) is  a formalism that relates a contact structure of  the
phase space $\mathcal{T}$ with the metric structure on a special
subspace of $\mathcal{T}$ called the space of equilibrium states
$\mathcal{E}$. In this work, we will use a Legendre invariant
metric in the context of geometrotermodynamics to formulate an
invariant geometric representation of the thermodynamics of a
static and spherically symmetric hairy black hole solution in
massive gravity.

This paper is organized as follows. In Section II, we review the
most important aspects of the generalized Schwarzschild solution
for massive gravity, emphasizing the thermodynamic interpretation
of its physical parameters. In Sec. III, we review the
fundamentals of GTD.  In Sec. IV, we investigate the GTD
 of a black hole in massive gravity and
show that it agrees with the results following from the analysis
of the corresponding thermodynamic variables. Finally, in Sec. V,
we present the conclusions of our work.

\section{Lorentz--breaking massive gravity}
\label{Fequation}

The general action that describes massive gravity is given by the
expression \cite{comelli,blas0,Rubakov,blas,capela,dubovsky2},

\bea \label{action}s=\int_{\mathcal{M}}d^4\sqrt{g}\Big[
-\frac{1}{16\pi}R+\Lambda^4 \mathcal{F}(X,V i,
W^{ij})\Big]-\int_{\partial
\mathcal{M}}d^3\sqrt{\gamma}\frac{K}{8\pi}\,,\eea where
$\mathcal{F}$ is a function of four scalar fields $\phi^\mu$ that
are minimally coupled to gravity and

\bea \label{action1}
X=\Lambda^{-4} g^{\mu \nu} \partial_\mu \phi^0 \partial_\nu \phi^0  \,\,; \,\, V^i=\Lambda^{-4}
\partial^\mu \phi^i \partial_\mu \phi^0 \,\,; \,\, W^{ij}=\Lambda^{-4}  \partial^\mu \phi^i \partial_\mu \phi^j-\frac{V^i V^j}{X}\,.\eea
Spacial and spacetime indices are denoted by latin and greek
letters, respectively. The second integral is the
Gibbons-Hawking-York boundary term \cite{bronne,comelli3}, where
$\gamma$ is the metric induced on the boundary $\partial
\mathcal{M}$ and $ K$ is the trace of the extrinsic curvature
$K_{ij}=\frac{1}{2}\gamma^k{}_i \nabla_k n_j $ of $\mathcal{M}$
with unit normal $n^i$. Such a boundary term is required to have a
well-defined variational principle in the presence of the border
$\partial \mathcal{M}$.

Spherically symmetric black holes in massive gravity have been
investigated in \cite{comelli,bronne,comelli3,bronne2} and it was
found that a set of coordinates always can be found where the
solution can be written in the form

\bea \label{metric} ds^2=f(r) dt^2+f(r)^{-1}dr^2+r^2(d\theta^2
\sin^2 \theta d\varphi^2)\,,\eea and the fields are given by the
expressions\cite{bronne}:

\bea \phi^0=\Lambda^2[t+h(r)]\,,\,\,\, ; \,\,\, \phi^i=\phi(r)
\frac{\Lambda^2 x^i}{r}\,.\eea

The metric functions $f$, $h$ and $\phi$ are given by the
following equations,

\bea \label{metricf} f&=& 1-\frac{2M}{r}-\frac{Q}{r^\lambda}\,\\
h&=&\pm \int \frac{dr}{f}\Bigg[ 1-f\Big(
\frac{\lambda(\lambda-1)Q}{12m^2 r^{\lambda+2}}
+1\Big)^{-1}\Bigg]^{\frac{1}{2}}\,,\\ \phi &=&r\,,\eea where $M$
and $Q$ are integration constants and $\lambda\neq1$  is a
positive constant. In the case $\lambda >1$ the gravitational
potential is asymptotically Newtonian and the parameter $M$
coincides with the ADM mass, while $Q$ is a scalar charge whose
presence reflects the modification of the gravitational
interaction as compared to General Relativity.

The roots of the lapse function $f$ ($g_{tt} = 0$) define the horizons $r = r_\pm$ of the spacetime. In particular, the null hypersurface $r = r_+$
can be shown to correspond to an event horizon, which in this case is also a Killing horizon,
whereas the inner horizon at $ r_-$ is a Cauchy horizon. Therefore from $f(r_+)=0$  \cite{capela} we get,

\bea \label{horizonte}
1-\frac{2M}{r_+}-\frac{Q}{r_+^\lambda}=0\,,\eea

From the expression (\ref{horizonte}) for the horizon
radii we obtain the following relationship,

\bea \label{funequ3} M(S,Q)=\frac{1}{2} r_+\Bigg[ 1-\frac{Q}{
r_+^{\lambda}}\Bigg]\,,\eea

From the area-entropy relationship, $S = \pi r_+^2$, the equation
(\ref{horizonte}) can be rewritten as

\bea \label{homoge} M(S,Q)=\frac{1}{2}\Bigg(\frac{S}{\pi}
\Bigg)^{\frac{1}{2}}\Bigg[ 1-\frac{Q}{\Big(\frac{S}{\pi}
\Big)^{\frac{\lambda}{2}}}\Bigg]\,.\eea

This equation relates all the thermodynamic variables entering the
black hole  metric in the form of a fundamental thermodynamic
equation $M = M(S, Q)$. Equation (\ref{homoge}) is an
inhomogeneous function in the extensive variables $S$ and $Q$.
Following Davies \cite{davies}, we homogenisize  the fundamental
equation by redefining the parameter $Q$ as,

\bea \label{chargeN} Q=q^{\frac{\lambda}{2}}\,.\eea Then, equation
(\ref{homoge}) becomes a homogeneous function of degree
$\frac{1}{2}$ in the extensive variables. This procedure was
performed explicitly in the context of GTD in \cite{Mustapha}. In
this case Euler's theorem takes the form \cite{davies},

\bea \label{EulerT1} \frac{1}{2}M=TS+\Theta
q^{\frac{\lambda}{2}}\,,\eea with

\bea \label{EulerT2} \Theta=\frac{2\Phi q}{\lambda
q^{\frac{\lambda}{2}} }=\frac{2\Phi
}{\lambda}q^{1-\frac{\lambda}{2}}\,.\eea

Differentiating equation  (\ref{EulerT1}) and using the first law
of thermodynamics we get,

\bea \label{EulerT3} \frac{1}{2}dM=dM-\Phi
dQ+SdT+\frac{\lambda}{2}\Theta
q^{\frac{\lambda}{2}-1}dQ+q^{\frac{\lambda}{2}}d\Theta\,.\eea Now,
we use (\ref{EulerT2}) in order to obtain,

\bea \label{EulerT4} dM=-2SdT-2q^{\frac{\lambda}{2}}d\Theta\,.\eea

The last relation implies the following thermodynamic equilibrium
conditions,

\bea \label{EulerT5} S&=&-\frac{1}{2}\frac{\partial M}{\partial
T}\,,\\ q^{\frac{\lambda}{2}}&=&-\frac{1}{2}\frac{\partial
M}{\partial \Theta}\,.\eea

According to the first law of the thermodynamics, the expression
for the temperature $T$ and the potential $\Theta$ are given by
the thermodynamic equilibrium conditions: $T =\Big( \frac{\partial
M}{\partial S}\Big)_{q}$ and $\Theta = \Big( \frac{\partial
M}{\partial q}\Big)_{S}$, which lead to the following results,

\bea \label{ThermCond2} T &=& \frac{1}{4\pi}\frac{\Big(
\frac{S}{\pi} \Big)^{\frac{\lambda}{2}}-
q^{\frac{\lambda}{2}}(1-\lambda)}{\Big(
\frac{S}{\pi} \Big)^{\frac{1+\lambda}{2}}}\,,\\
\label{ThermCond3}\Theta &=& -\frac{\lambda}{4}\Big( \frac{S}{\pi}
\Big)^{\frac{1-\lambda}{2}}  q^{\frac{\lambda}{2}-1}\,.\eea

It is easy to show that the temperature (\ref{ThermCond2})
coincides with the Hawking temperature. The temperature $T$ will
be positive when $ \Big(\frac{S}{\pi} \Big)^{\frac{\lambda}{2}}-
q^{\frac{\lambda}{2}}(1-\lambda)>0$, i.e., if $\Big(\frac{S}{\pi}
\Big)^{\frac{\lambda}{2}}>q^{\frac{\lambda}{2}}(1-\lambda)$. The
temperature increases rapidly as a function of entropy $S$ until
it reaches its maximum value at $\Big(\frac{S}{\pi}
\Big)^{\frac{\lambda}{2}}=q^{\frac{\lambda}{2}}(1-\lambda^{2})$.
Then, as the entropy increases, the temperature becomes a
monotonically decreasing function. The behavior is shown in fig.1

\begin{figure}[h]
{\includegraphics[scale=0.4]{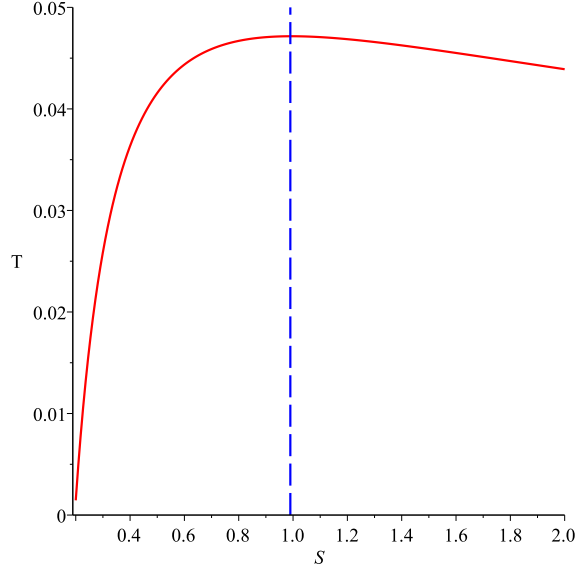} \caption{The
temperature $T$ as a function of the entropy $S$, with
$\lambda=\frac{1}{2}$ and $q=1$.}}\end{figure}

The heat capacity  at constant values of $q$ is given as

\bea \label{capacity} C_{q}=T\Bigg(\frac{\partial S}{\partial
T}\Bigg)_{q}=\Bigg(\frac{\frac{\partial M}{\partial
S}}{\frac{\partial^2 M}{\partial S^2}}\Bigg)_{q}\,,\eea where the
subscript indicates that derivatives are calculated keeping the
charge constant. Using the fundamental equation (\ref{homoge}) we
get,

\bea \label{heatcapacity}C_{q}=-\frac{2S\Big[\Big(\frac{ S}{\pi}
\Big)^{\frac{\lambda}{2}}+q^{\frac{\lambda}{2}}(\lambda-1)
\Big]}{\Big[\Big(\frac{ S}{\pi}
\Big)^{\frac{\lambda}{2}}-q^{\frac{\lambda}{2}}(1-\lambda^2)
\Big]}\,.\eea

According to Davies \cite{davies}, second order phase transitions
take place at those points where the heat capacity diverges, i.
e., for

\bea \label{sing} \Big(\frac{ S}{\pi}
\Big)^{\frac{\lambda}{2}}-q^{\frac{\lambda}{2}}(1-\lambda^2)=0\,\eea
these points exist in the interval $1-\lambda^2>0$. This behavior
is depicted in figure 2.

\begin{figure}[h]
{\includegraphics[scale=0.4]{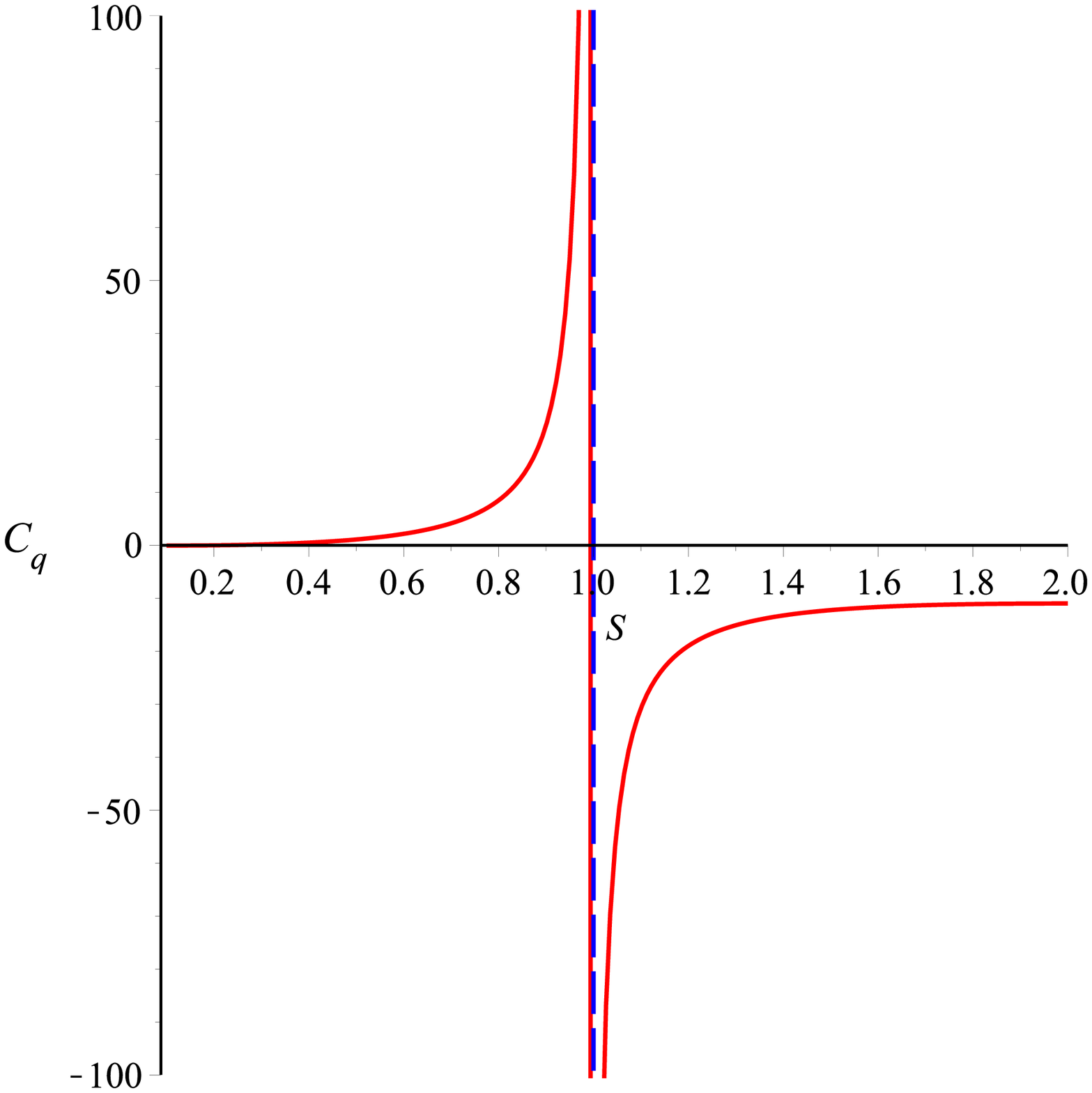}\includegraphics[scale=0.4]{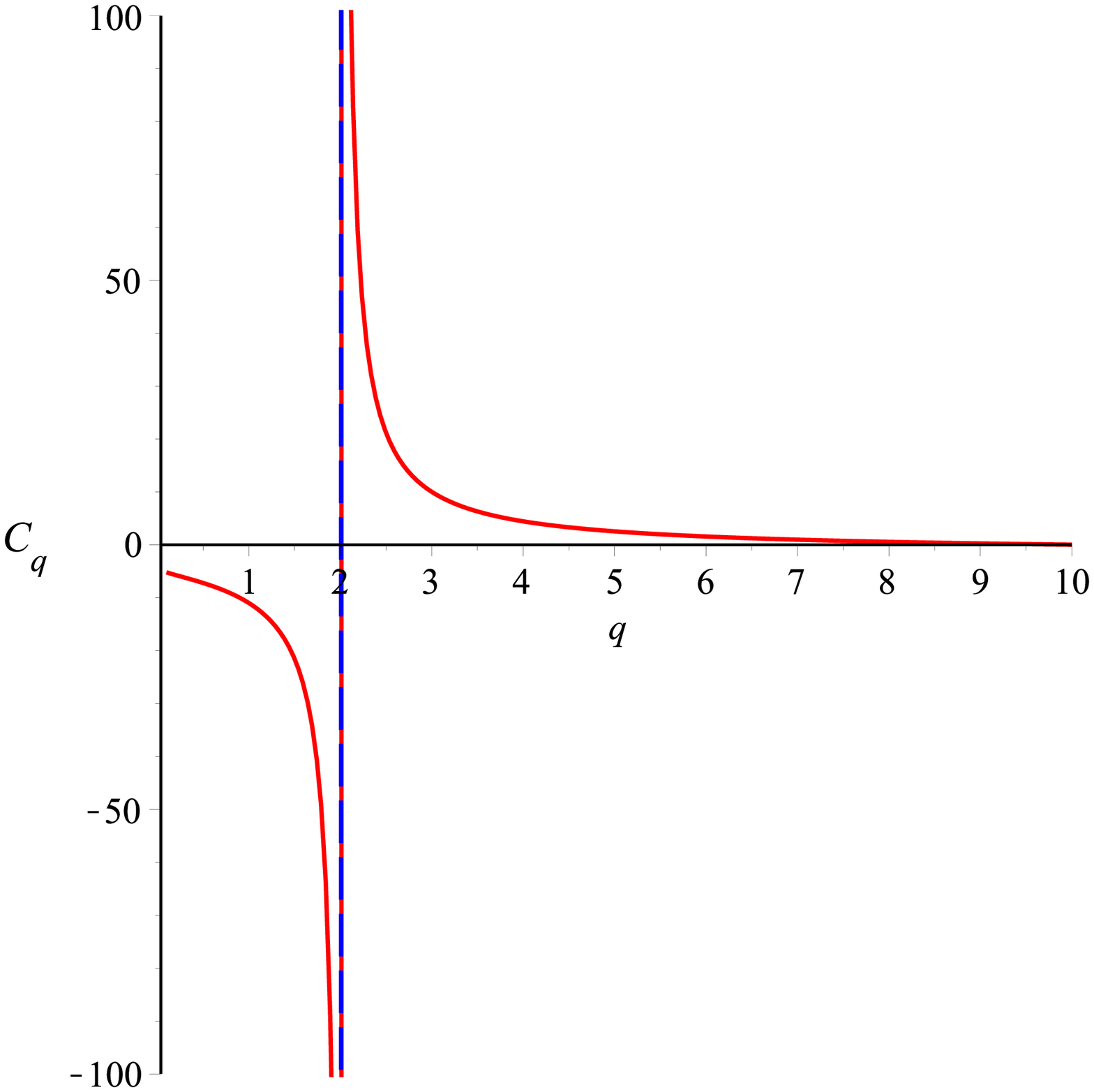}
\caption{The heat capacity $C_{q}$ as a function of the entropy
$S$ (left) with $q=1$ and as a function of the charge $q$ (right)
with $S=2$. In both case we have chosen $\lambda=\frac{1}{2}$,
}}\end{figure}

In the physical region with $ \Big(\frac{S}{\pi}
\Big)^{\frac{\lambda}{2}}- q^{\frac{\lambda}{2}}(1-\lambda)> 0$.
i.e., the region with positive temperature, the heat capacity is
positive in the interval,

\begin{equation}
q^{\frac{\lambda}{2}}(1-\lambda)<\Big(\frac{S}{\pi}
\Big)^{\frac{\lambda}{2}}<q^{\frac{\lambda}{2}}(1-\lambda^2)\,,
\end{equation}indicating that the black hole is stable in this
region. At the maximum value of the temperature which occurs at $
\Big(\frac{S}{\pi} \Big)^{\frac{\lambda}{2}}-
q^{\frac{\lambda}{2}}(1-\lambda^2)$, the heat capacity diverges
and changes spontaneously its sign from positive to negative. This
indicates the presence at second order phase transition which is
accompanied by a transition into a region of instability.

\section{ Review of Geometrothermodynamics} \label{review}

Geometrothermodynamics is a theory that has been formulated in
order to introduce the Lagendre invariance in the geometric
description of the thermodynamic equilibrium states
\cite{quevedo2}. This theory has been applied to different
thermodynamic systems like black holes, ideal gas or Van der Waals
gas \cite{quevedo,quevedo3,quevedo4,
quevedo5,quevedo6,quevedo7,quevedo8,quevedo9}. In all the cases
investigated so far, GTD leads to consistent results that describe
geometrically the phases transitions and thermodynamic interaction
using Legendre invariant metrics.

The main ingredient of  GTD is a $(2n+1)$--dimensional manifold
$\mathcal{T}$ with a set of coordinates $Z^A$ which allow us to
define a non--degenerate Legendre invariant metric $G$ together
with a linear differential 1--form $\Theta$ which fulfills the
condition $\Theta \wedge (d\Theta)^n \neq 0$, where $n$ is the
number of thermodynamic degrees of freedom, $\wedge$ represents
the exterior product and $d$ the exterior derivative. In GTD, we
also have  the space of thermodynamic equilibrium states which is
a submanifold $\mathcal{E}\subset \mathcal{T}$  defined by means
of a smooth embedding mapping $\varphi :\mathcal{E}\longrightarrow
\mathcal{T}$ such that the pullback $\varphi^*(\Theta)=0$.

With the above elements a metric $g$  is induced in $\mathcal{E}$
by means of $\varphi^*(G)=g$, giving a Riemannian structure to
this space. So, in GTD the physical properties of a thermodynamic
system in a state of equilibrium  are described in terms of the
geometric properties of the corresponding space $\mathcal{E}$.

The set of elements $(\mathcal{T}, \Theta, G)$ with the conditions
above mentioned is called a contact Riemannian manifold. If we
consider the $2n+1$--dimensional space $\mathcal{T}$ coordinatized
by the set $Z^A=\Big\{\Phi\,,E^a\,,I^a \Big\}$ where
$A=0,\dots,2n$ and $a=1,\dots n,$ the 1-form $\Theta$ will be

\bea \label{gtd1} \Theta=d\Phi-I_adE^a\,.\eea

We choose now the subset $E^a$ as coordinates of $\mathcal{E}$.
Then, the mapping $\varphi$ is given by

\bea \label{gtd2} \varphi: (E^a) \longrightarrow
(\Phi\,,E^a\,,I^a)\,,\eea and the condition,

\bea \label{gtd3}
\varphi^*(\Theta)=\varphi^*(d\Phi-\delta_{ab}I^adE^b)=0\,,\eea
leads to the standard conditions of the thermodynamic equilibrium
and the first law of thermodynamics,

\bea \label{gtd4} \frac{\partial \Phi}{\partial E^a}=I_a\,, \quad
\quad \quad \quad d\Phi= I_adE^a \,.\eea The second law of
thermodynamics under this formalism is written as,

\bea \label{gtd5} \frac{\partial^2 U}{\partial E^a \partial E^b
}\geq 0\,, \quad \quad ; \quad \quad \frac{\partial^2 S}{\partial
F^a \partial F^b }\leq 0 \,,\eea where $U$ and $S$ represent the
energy and entropy for each of the corresponding thermodynamic
systems. Here $E^a$ ($F^a$) represent all the extensive variables
other than $U$ ($S$).

In GTD the only  requirement for defining a metric $G$ of the
space $\mathcal{T}$ is that it fulfills the condition of Legendre
invariance; therefore, we have many possibilities of constructing
a metric with these features. In fact, all the Legendre invariant
metrics found so far can be classified in three classes. It turns
out that each class can be used to describe thermodynamic systems
with particular phase transitions \cite{quevedo3}. For instance,
for systems with second order phase transitions, the most general
metric can be expressed as,

\bea \label{gtd6} G=\Theta^2+(\delta_{ab}E^a I^b)(\eta_{cd}dE^c
dI^d)\,,\eea where $\delta_{ab}={\mathrm{diag}}(1,1,\dots,1)$ and
$\eta_{ab}={\mathrm{diag}}(-1,1,\dots,1)$ . It can be shown that
the metric (\ref{gtd6}) is invariant with respect to a total
Legendre transformation which changes the coordinates $\{\Phi,
E^a, I^a\}$ to the coordinates $\{\tilde{\Phi},
\tilde{E^a},\tilde{I^a}\}$ using the following algebraic rules,

\bea \label{gtd7} \Phi=\tilde{\Phi}-\tilde{E}_a
\tilde{I}^a\,,\quad \quad E^a=-\tilde{I}^a\,, \quad \quad
I^a=\tilde{E}^a\,. \eea Applying the pullback $\varphi^*$ to the
metric (\ref{gtd6}), we obtain the corresponding thermodynamic
metric $g$,

\bea \label{gtd8} g^{GTD}=\varphi^*(G)=\Big( E^c \frac{\partial
\Phi}{\partial E^c} \Big)\Big(\eta_{ab}
\delta^{bc}\frac{\partial^2 \Phi}{\partial E^c
\partial E^d }dE^a dE^d\Big)\,,\eea which depends only of the
fundamental potential $\Phi=\Phi(E^a)$. If we know the fundamental
potential of the thermodynamic system that we want to study,  the
corresponding metric $g$ can be computed explicitly and the
relations between thermodynamic and geometry can be also studied.

\section{Geometrothermodynamics in massive gravity} \label{Fequation}

Let us consider a thermodynamic system with two degrees of
freedom. If we choose thermodynamic potential as $\Phi=M$ and the
coordinates of equilibrium manifold as $E^a=\{ S, q \}$, then the
corresponding metric is given as

\bea \label{GTDmetric1} g^{GTD}=\Big( S \frac{\partial M}{\partial
S} + q \frac{\partial M}{\partial q}\Big)\Big(-\frac{\partial^2
M}{\partial S^2 }dS^2 +\frac{\partial^2 M}{\partial q^2 }dq^2\Big)
\,.\eea

Using the expressions for the thermodynamic $M$, as given in Eq.
(\ref{homoge}), we obtain explicitly the GTD metric coefficients,
which can be written as,

\bea \label{comp11}
g^{GTD}_{SS}&=&\frac{1}{32\pi^2}\frac{\Big[\Big(\frac{ S}{\pi}
\Big)^{\frac{\lambda}{2}}-q^{\frac{\lambda}{2}}
\Big]\Big[\Big(\frac{ S}{\pi}
\Big)^{\frac{\lambda}{2}}-q^{\frac{\lambda}{2}}(1-\lambda^2)
\Big]}{\Big(\frac{ S}{\pi} \Big)^{\lambda+1}}\,,\eea

\bea \label{comp22} g^{GTD}_{qq}&=&-\frac{1}{32}\Big(\frac{
S}{\pi} \Big)^{1-\lambda}q^{\frac{\lambda}{2}-2}\Big[\Big(\frac{
S}{\pi} \Big)^{\frac{\lambda}{2}}-q^{\frac{\lambda}{2}}
\Big]\lambda(\lambda-2)\,,\eea and Eq. (\ref{GTDmetric1}) takes
the form

\bea \label{GTDmetric2}
g^{GTD}=\frac{1}{32\pi^2}\frac{\Big[\Big(\frac{ S}{\pi}
\Big)^{\frac{\lambda}{2}}-q^{\frac{\lambda}{2}} \Big]}{\Big(\frac{
S}{\pi} \Big)^{\lambda+1}} \Bigg\{\Big[\Big(\frac{ S}{\pi}
\Big)^{\frac{\lambda}{2}}+q^{\frac{\lambda}{2}}(\lambda^2-1)
\Big]dS^2-\lambda (\lambda-2)\Big(\frac{S}{q}\Big)^2
q^{\frac{\lambda}{2}}dq^2 \Bigg\}\,.\eea

The curvature scalar corresponding to the metric
(\ref{GTDmetric1}) takes the form,

\bea \label{scalar} R^{GTD}=-\frac{8\pi \lambda \Big(\frac{
S}{\pi}
\Big)^{\frac{3\lambda}{2}}\Bigg[q^{\frac{\lambda}{2}}(2\lambda^2+2\lambda-3)+\Big(\frac{
S}{\pi} \Big)^{\frac{\lambda}{2}} (3-2\lambda)\Bigg]\Bigg[
\Big(\frac{ S}{\pi}
\Big)^{\frac{\lambda}{2}}+(\lambda-1)q^{\frac{\lambda}{2}}\Bigg]}{S(\lambda-2)\Big[\Big(\frac{
S}{\pi}
\Big)^{\frac{\lambda}{2}}-q^{\frac{\lambda}{2}}(1-\lambda^2)
\Big]^2 \Big[ q^{\frac{\lambda}{2}}-\Big(\frac{ S}{\pi}
\Big)^{\frac{\lambda}{2}}\Big]^3}\,, \eea

There are two curvature singularities in this case. The first one
occurs if $q^{\frac{\lambda}{2}}-\Big(\frac{ S}{\pi}
\Big)^{\frac{\lambda}{2}}=0 $ and corresponds to $ M = 0$, as
follows from Eq.(\ref{homoge}), see also figure 3. This means that
this singularity is non physical since no black hole is present in
this case. A second singularity is located at the roots of the
equation $\Big(\frac{ S}{\pi}
\Big)^{\frac{\lambda}{2}}-q^{\frac{\lambda}{2}}(1-\lambda^2)= 0$.
For the interval $1-\lambda^2>0$, according to
(\ref{heatcapacity}), it coincides with the points where
$C\longrightarrow \infty$, i. e., with the points where second
order phase transitions take place. The general behavior of the
curvature scalar is illustrated in figures 4.

\begin{figure}[h]
{\includegraphics[scale=0.4]{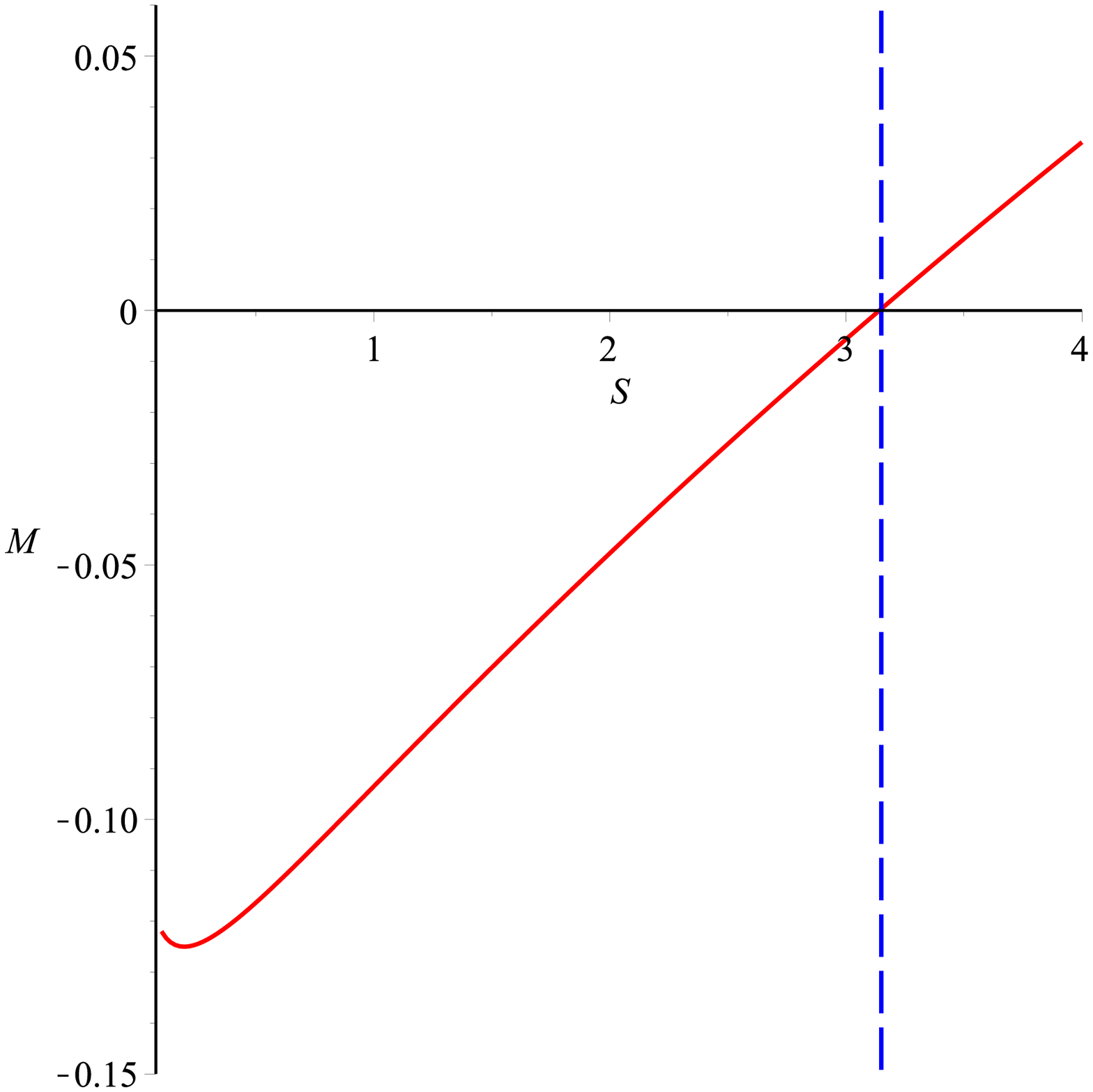}\includegraphics[scale=0.4]{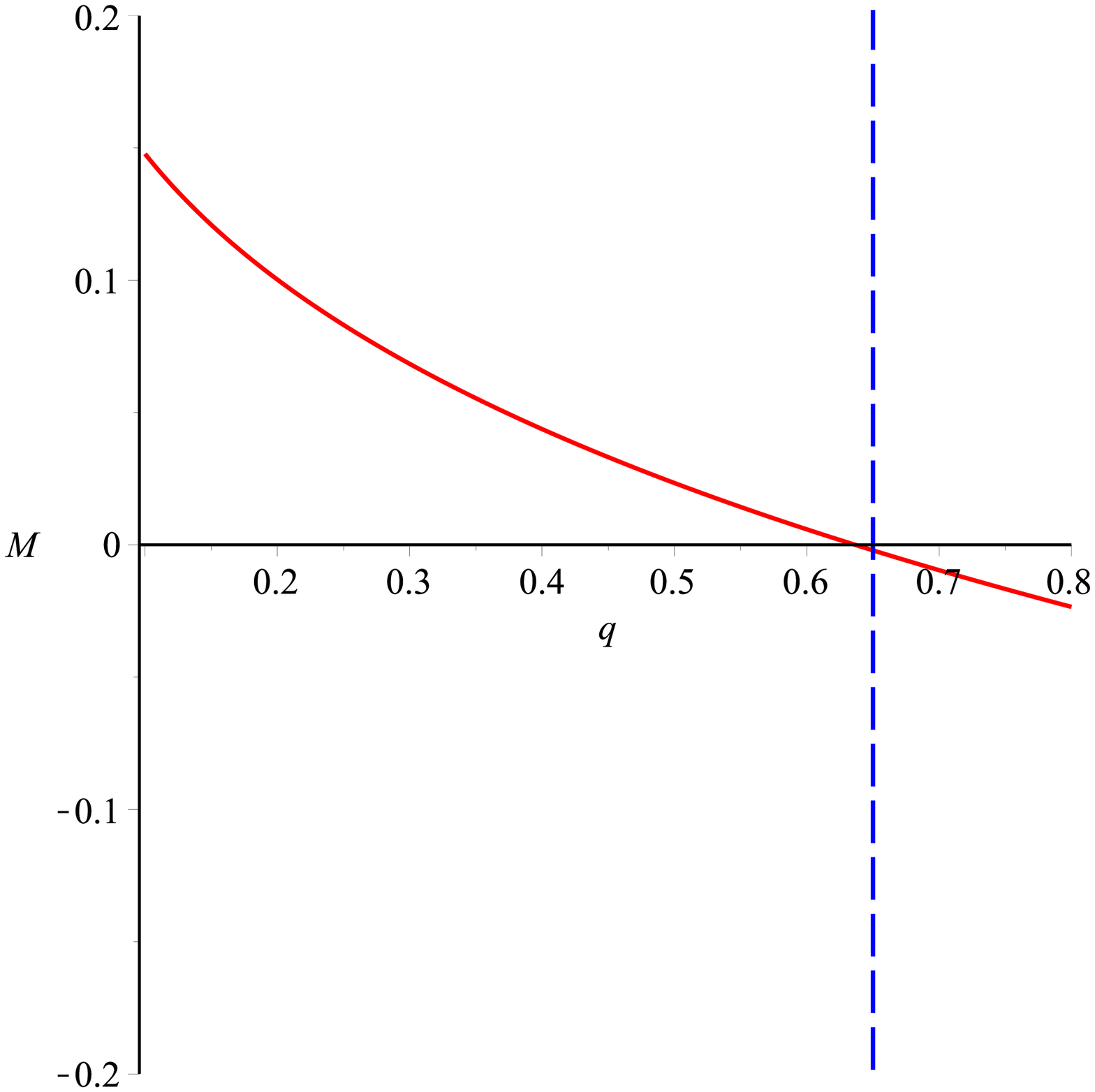}
\caption{The mass $M$ as a function of the entropy $S$ (left) and
as a function of charge $q$ right, with $q=1$ and $S=2$,
respectively. We have considered $\lambda=\frac{1}{2}$ in both
cases.}}\end{figure}

\begin{figure}[h]
{\includegraphics[scale=0.4]{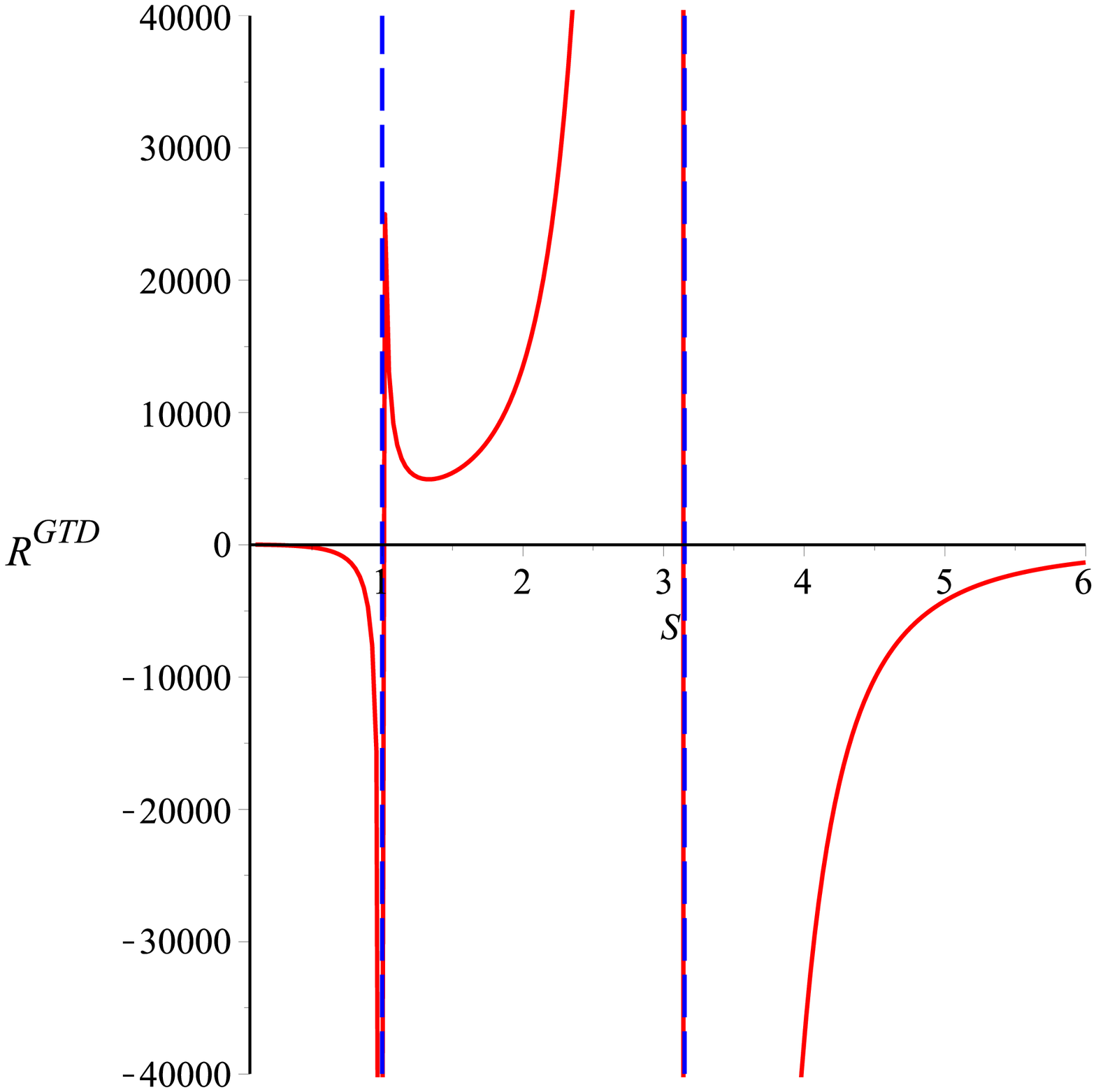}\includegraphics[scale=0.4]{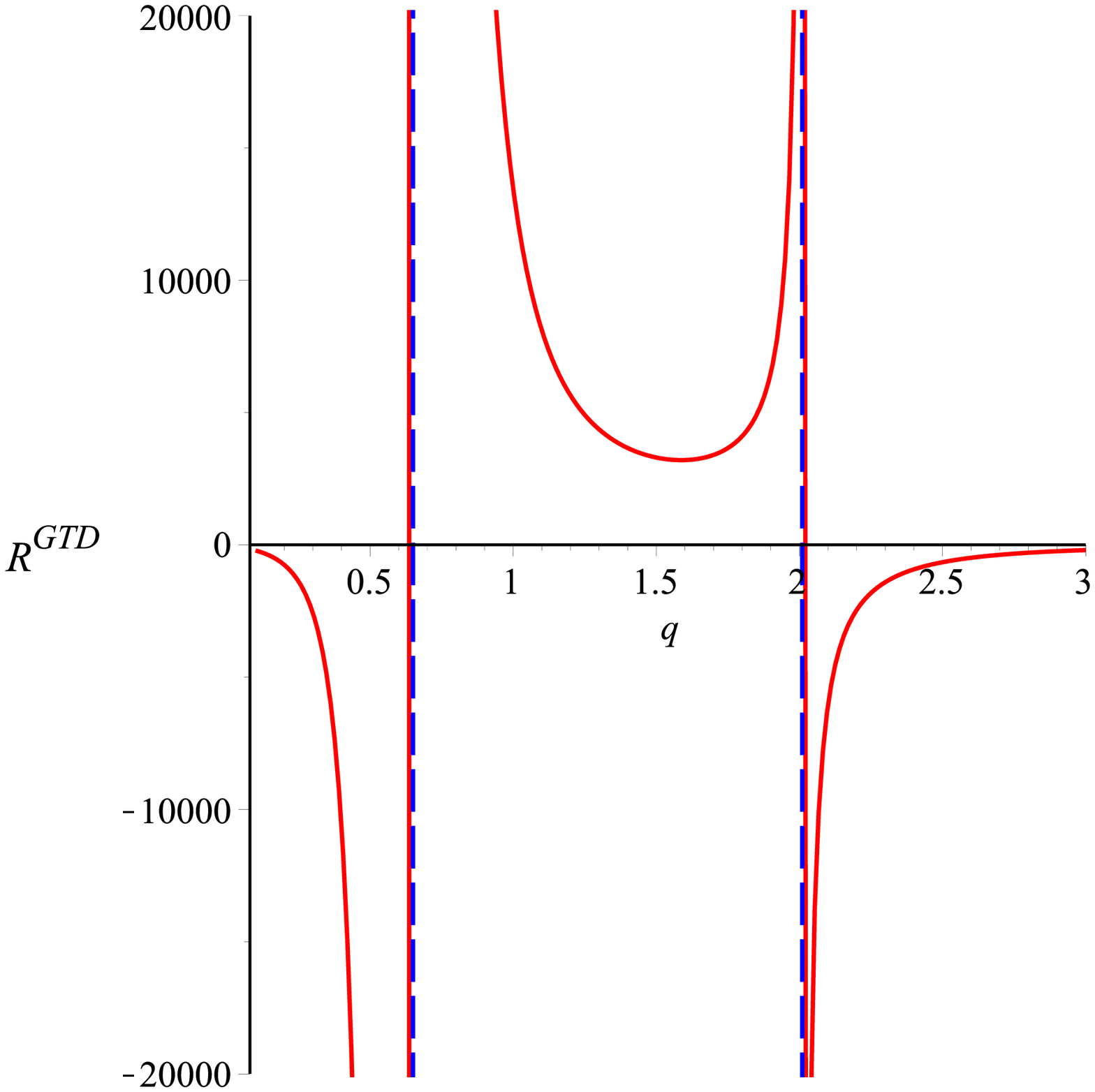}
\caption{The curvature scalar $R^{GTD} $ as a function of the
 entropy (left), with $q=1$, and as a function of charge $q$ (right), with $S=2$. We have considered $\lambda=\frac{1}{2}$ in both cases.}}\end{figure}
\newpage

According to GTD, these results show that there exist curvature
singularities at those points where second order phase transitions
occur, because the denominators of the heat capacity and the
 curvature scalar coincide \cite{quevedo3}. We can also observe,
 figure 4, that the curvature scalar changes spontaneously its sign
 indicating a transition of the a stable region to the unstable
 one, reproducing the thermodynamic behavior of this black hole.

\section{Weinhold and Ruppeiner Approaches} \label{Weinhold}

In this section we analyze the thermodynamic geometry of the same
black hole in massive gravity by using Weinhold and Ruppeiner
metrics.

The Weinhold metric is defined as \cite{Weinhold}

\bea \label{Weinholdmetric} g^{W}=(\frac{\partial^2 M}{\partial
S^2 }dS^2 +2\frac{\partial^2 M}{\partial S \partial q
}dS^2+\frac{\partial^2 M}{\partial q^2 }dq^2\Big) \,.\eea

Using the fundamental thermodynamic equation $M = M(S, q)$, the
metric (\ref{Weinholdmetric}) can be calculated and we obtain
explicitly the Weinhold metric coefficients which can be written
as,

\bea \label{Weinholmetric2} g^{W}=-\frac{1}{8\pi^2\Big(\frac{
S}{\pi} \Big)^{\frac{3}{2}(\lambda+3)}} \Bigg\{&&\Big[\Big(\frac{
S}{\pi}
\Big)^{\frac{\lambda}{2}}+q^{\frac{\lambda}{2}}(\lambda^2-1)
\Big]dS^2-Sq^{\frac{\lambda}{2}-1}\lambda(\lambda-1)dSdq+\nonumber
\\ &+&S^2   q^{\frac{\lambda}{2}-2}\lambda(\lambda-2)
dq^2 \Bigg\}\,.\eea

The curvature scalar corresponding to the metric
(\ref{Weinholmetric2}) takes the form,

\bea \label{scalarW} R^{W}=0\,, \eea

This result tells us that the space of thermodynamic equilibrium
is flat, indicating the lack of thermodynamic interaction.
Obviously, this result contradicts the results obtained by using
black hole thermodynamics.

The case of Ruppeiner's geometry for thermodynamical systems must
be computed in the entropy representation. However, as has been
shown in \cite{Mrugala} one can prove that Ruppeiner's metric,
$g^R$, is proportional to Weinhold's metric, $g^W$ as $g^R =
(1/T)g^W$, where $T$ is the temperature. Using this result
Ruppeiner's metric takes the form,

\bea \label{Ruppeiner2}
g^{R}=-\frac{1}{2S\Big[q^{\frac{\lambda}{2}}(\lambda-1)+\Big(\frac{
S}{\pi} \Big)^{\frac{\lambda}{2}}\Big]} \Bigg\{&&\Big[\Big(\frac{
S}{\pi}
\Big)^{\frac{\lambda}{2}}+q^{\frac{\lambda}{2}}(\lambda^2-1)
\Big]dS^2-Sq^{\frac{\lambda}{2}-1}\lambda(\lambda-1)dSdq+\nonumber
\\ &+&S^2   q^{\frac{\lambda}{2}-2}\lambda(\lambda-2)
dq^2 \Bigg\}\,.\eea The corresponding curvature scalar can be
written as

\bea \label{scalar} R^{Rup}=\frac{\lambda \Big(\frac{ S}{\pi}
\Big)^{\frac{\lambda}{2}}\Bigg[q^{\frac{\lambda}{2}}(\lambda^2+\lambda-2)-\Big(\frac{
S}{\pi} \Big)^{\frac{\lambda}{2}}
(\lambda-2)\Bigg]}{S\Big[\Big(\frac{ S}{\pi}
\Big)^{\frac{\lambda}{2}}+q^{\frac{\lambda}{2}}(\lambda-1) \Big]
\Big[ 2q^{\frac{\lambda}{2}}(\lambda-1)-\Big(\frac{ S}{\pi}
\Big)^{\frac{\lambda}{2}}(\lambda-2)\Big]}\,, \eea

\begin{figure}[h]
{\includegraphics[scale=0.4]{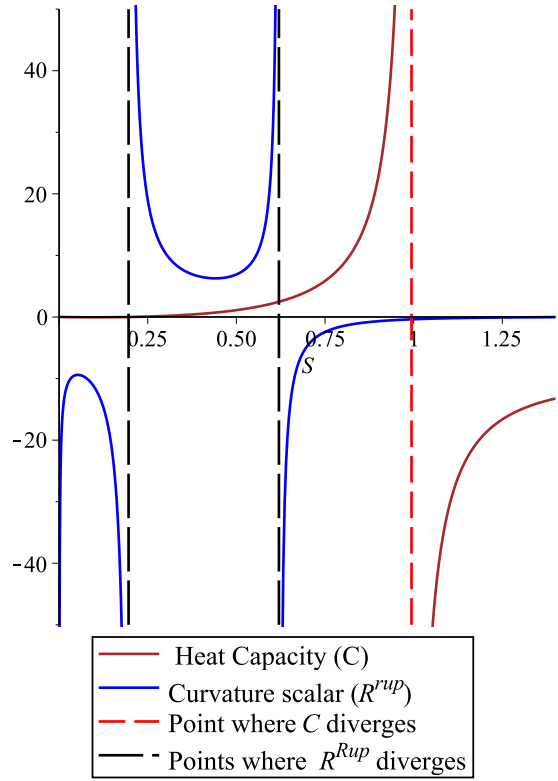} \caption{The curvature
scalar $R^{Rup}$ (in blue) and heat capacity (in brown) as
functions of $S$, with $q = 1$ and $\lambda = \frac{1}{2}$. The
red dash  thick vertical line corresponds to the point where the
heat capacity is singular and the black longdash  vertical lines
represent the points where the curvature scalar of the Ruppeiner
metric diverges.}}\end{figure}
\newpage

We can see that the Ruppeiner's geometry is curved, signaling
interaction for this thermodynamic system. There are singular
points, but they are not consistent with the divergencies of the
heat capacity for fixed charge.

The fact that Ruppeiner's curvature scalar in this case doesn't
diverge at the same points where the heat capacity does, tell us
that it is not possible to associate curvature singularities with
second-order phase transitions.

\section{Conclusions} \label{conclusions}

In this work, we analyzed  the thermodynamic and
geometrothermodynamic properties of a black hole solution in
massive gravity. Using the fundamental equation of a spherically
symmetric black hole we found the geometric properties of the
corresponding manifold of equilibrium states. We found that the
corresponding thermodynamic curvature turned out to be nonzero,
indicating the presence of thermodynamic interaction. A numerical
and analytical study of the thermodynamic curvature shows that the
phase transitions which are characterized by divergencies of the
heat capacity are described in GTD by curvature singularities in
the equilibrium manifold.

We assumed in this work Davies's proposal to solve the  problem of
the lack of homogeneity of the fundamental equation. We have
considered the definition (\ref{chargeN}), with which the
fundamental equation (\ref{funequ3}) becomes a homogeneous
function of degree $\frac{1}{2}$  in the extensive variables. We
have also shown that for values of the constant $\lambda$ in the
interval $\lambda^2-1 < 0$, the black holes in massive gravity
presents a phase transition. Therefore, it would be interesting to
investigate further  $\lambda$ as a thermodynamic variable in
order to learn more about the thermodynamics and phase transition
structure of these black holes.

We analyzed the thermodynamic geometry based on the Weinhold
metric and found that it represents a flat space, indicating the
lack of thermodynamic interaction. We also analyzed the
Ruppeiner's metric for this system and found that corresponds to a
curved manifold for this black hole and the corresponding
curvature diverges at some points, but these points are not the
ones at where the heat capacity for fixed charge diverges. We
conclude that the Weinhold and Ruppeiner's geometries do not
describe correctly the thermodynamic geometry of black holes in
massive gravity. We interpret this result as indication that it is
necessary to take into account Legendre invariance in order to
correctly describe thermodynamics from a geometric point of view.

\section*{Acknowledgements}

This work was supported by Conacyt-Mexico, Grant No. 166391 and
DGAPA-UNAM, Grant No.113514. I am grateful to Profs. F. Nettel and
C. Lopez-Monsalvo for interesting comments, encouragement and
support.


\begin{thebibliography}{99}



\bibitem{Donoghue}J. F. Donoghue, {\em Introduction to the Effective Field Theory Description of Gravity}, arXiv:gr-qc/9512024.

\bibitem{Copeland} E. J. Copeland, M. Sami and S. Tsujikawa, Int. J. Mod. Phys. D {\bf 15}, 1753–-1936 (2006).

\bibitem{Arkani} N. Arkani-Hamed, H.Georgi and  M.D. Schwartz, Annals Phys., {\bf 305}, 96--118 (2003).

\bibitem{Arkani2} N. Arkani-Hamed, H.C Cheng, M.A. Luty and S. Mukohyama, JHEP,{\bf 0405}, 074 (2004) .

\bibitem{dubovsky} S.L. Dubovsky, JHEP, {\bf 0410}, 076 (2004).


\bibitem{comelli} D. Comelli, F. Nesti and L. Pilo, Phys. Rev. D {\bf 83}, 084042(2011).

\bibitem{Bravetti} A. Bravetti, D. Momeni, R. Myrzakulov and H. Quevedo, Gen. Rel. Grav.
{\bf 45},1603-1617 (2013).

\bibitem{Aman}J. E. {\AA}man, I. Bengtsson, and N. Pidokrajt, Gen. Rel. Grav. {\bf
35}, 1733 (2003).

\bibitem{Aman2} J. E. {\AA}man and N. Pidokrajt, Phys. Rev. D {\bf 73}, 024017 (2006).

\bibitem{Aman3} J. E. {\AA}man and N. Pidokrajt, Gen. Rel. Grav. {\bf 38}, 1305 (2006).

\bibitem{shen} J. Shen, R. G. Cai, B. Wang, and R. K. Su, Int. J. Mod. Phys. A {\bf 22}, 11--27 (2007).

\bibitem{Cai}  R. G. Cai and J. H. Cho, Phys. Rev. D {\bf 60}, 067502 (1999); T.
Sarkar, G. Sengupta, and B. N. Tiwari, J. High Energy Phys. {\bf
0611} 015 (2006).

\bibitem{Weinhold} F. Weinhold, J. Chem. Phys. {\bf 63},2488 (1975).

\bibitem{Ruppeiner} G. Ruppeiner, Phys. Rev. A {\bf 20}, 1608 (1979).

\bibitem{quevedo2} H. Quevedo, J. Math. Phys. {\bf 48}, 013506 (2007).

\bibitem{blas0}D. Blas, {\em Aspects of Infrared Modifications of Gravity}, arXiv:0809.3744v1 [hep-th].

\bibitem{Rubakov} V. A. Rubakov,{\em  Lorentz-violating graviton masses: getting around ghosts, low strong coupling
scale and VDVZ discontinuity}, hep-th/0407104, (2004).

\bibitem{blas}D. Blas, D. Commelli, F. Nesti and L. Pilo, Phys. Rev. D, {\bf 80}, 044025 (2009).

\bibitem{capela} Capela F. and Nardini G, Phys.Rev. D {\bf 86}, 024030 (2012).


\bibitem{dubovsky2} S.L. Dubovsky, P. Tinyakov and M. Zaldarriaga, JHEP, {\bf 0711}, 083
(2007).

\bibitem{bronne} M. V. Bebronne and P. G. Tinyakov,  JHEP {\bf 0904}, 100 (2009)
[Erratum-ibid. 1106 (2011) 018].

\bibitem{comelli3} Z. Berezhiani, D. Commelli, F. Nesti and L. Pilo, JHEP {\bf 0807}, 130 (2008).

\bibitem{bronne2} M. V. Bebronne, {\em Theoretical and phenomenological aspects of theories with massive gravitons},
arXiv:0910.4066 [gr-qc].

\bibitem{davies} P.C.W Davies,Proc. Roy. Soc. Lond. A,
{\bf 353} 499 (1977).

\bibitem{Mustapha} Mustapha Azreg-Ainou, European Physical Journal C {\bf 74}, 1--8 (2014).

\bibitem{quevedo} H. Quevedo, Gen. Rel. Grav. {\bf 40}, 971 (2008).

\bibitem{quevedo3}H. Quevedo, A. S\'anchez, S. Taj, and A. V\'azquez, Gen.Rel.Grav. {\bf 43}, 1153-1165 (2011)..


\bibitem{Ruppeiner2}G. Ruppeiner, Rev. Mod. Phys. {\bf 67}, 605 (1995); {\bf 68}, 313 (1996).


\bibitem{Mrugala}R. Mrugala, Rep. Math. Phys. {\bf 21} 197--203(1985).


\bibitem{quevedo4} J. L. \'Alvarez, H. Quevedo, and A. S\'anchez, Phys. Rev. D {\bf 77}, 084004 (2008).

\bibitem{quevedo5} A. V\'azquez, H. Quevedo, and A. S\'anchez, J. Geom. Phys. {\bf 60}, 1942 (2010).

\bibitem{quevedo6} H. Quevedo, A. S\'anchez and A. V\'azquez,  Gen.Rel.Grav. {\bf 47} 4,  36 (2015).

\bibitem{quevedo7}H. Quevedo and A. S\'anchez, JHEP {\bf 09}, 034 (2008).

\bibitem{quevedo8} H. Quevedo and A. S\'anchez, Phys. Rev. D {\bf 79}, 024012 (2009).

\bibitem{quevedo9} H. Quevedo and A. S\'anchez, Phys. Rev. D. {\bf 79}, 087504 (2009).

\bibitem{Weinberg} S. Weinberg. {\em The cosmological constant problems},
arXiv:astro-ph/0005265v1.

\bibitem{Weinberg2} S. Weinberg, Rev. Mod. Phys. {\bf 61}, 1–-23 (1989).

\bibitem{Ruppeiner2} G. Ruppeiner, Rev. Mod. Phys. {\bf 67}, 605 (1995); {\bf 68}, 313 (1996).

\bibitem{Medved} A. J. M. Medved, Mod. Phys. Lett. A {\bf 23}, 2149 (2008).



\end{thebibliography}
\end{document}